\documentclass[12pt]{article} 
\usepackage{psfig} 
 
  \advance\oddsidemargin by 15mm 
  \advance\evensidemargin by 15mm 
  \advance\topmargin by 16mm 
 
\begin{document} 
 
\noindent 
\begin{minipage}[t][2cm][t]{18cm} 
 
{\large\bf   Angle-resolved  study of density-waves, 
superconductivity and pseudogap
in two dimensions  
}\\ 
 
\smallskip 
 
Dra\v{z}en Zanchi \\ 
{\it Laboratoire de Physique Th\'eorique et Hautes Energies, 
Paris }\\

\end{minipage}

\begin{minipage}[t][1cm][t]{18cm} 

\end{minipage} 

\noindent 
\hspace*{15mm}\parbox[t]{16.5cm} {\small 
 
  {\bf Abstract.} Weakly correlated electrons on a square lattice 
are studied by
angle-resolved functional renormalization group. Upon renormalization 
the interaction starts to depend on momenta and has pole-like 
solutions near a doping-dependent characteristic critical energy scale.
Near  half-filling this scale is the pseudogap temperature
$T^*$. In the overdoped regime the critical scale is the mean-field like
critical temperature for d-wave superconductivity. } 
 
\begin{figure} 
\centerline{\psfig{file=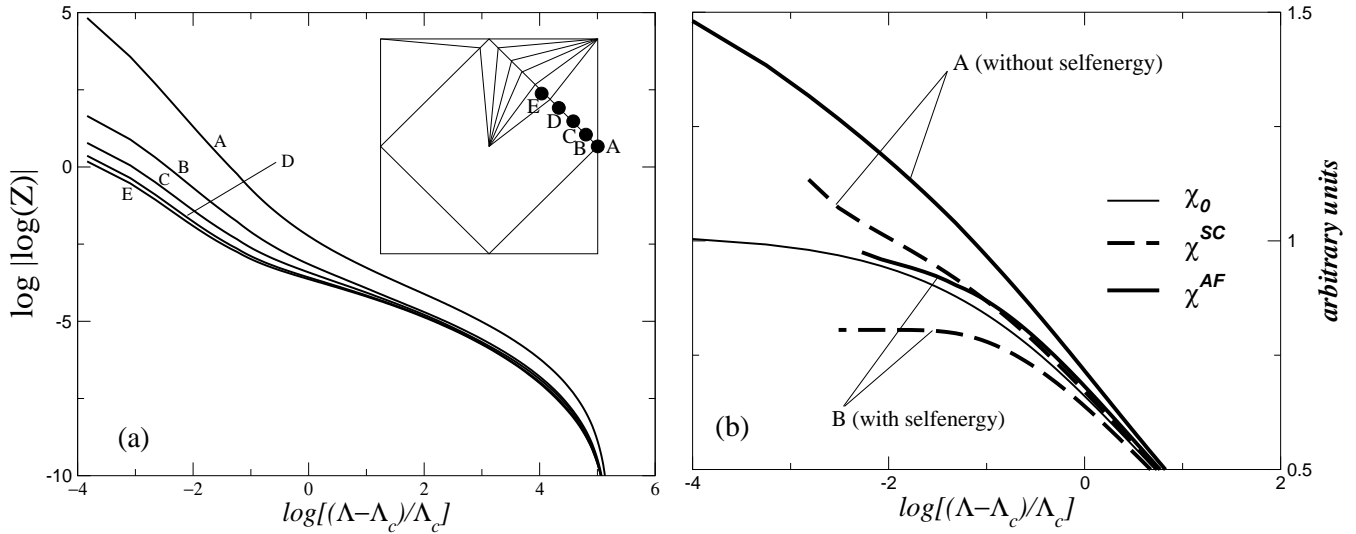,width=0.99\textwidth,angle=0} }
\hfill 
 
\caption{\small 
 (a)  Flow of  the angle-resolved quasiparticle weight on several points at 
the Fermi
surface.Inset  shows the Brillouin zone with 32-patch 
discretization and with the
square Fermi surface.   (b) The relevant susceptibilities for 
d-wave superconductivity (dashed line) and for
antiferromagnetism (solid line) without (A) and with (B) selfenergy insertions. Thin line
represents the bare susceptibility. 
  } 
\label{results} 
\end{figure}

\vspace{0.5cm} 
 I would like to explain in which precise points the angle 
resolved one-loop  renormalization group analysis of the 
effective action for Landau quasiparticles 
can increase our understanding of low temperature properties of 
interacting fermions on a lattice.
Our attention will be focused on the underdoped  high-$T_c$
superconductors as a prototype of correlated electronic 
system with strong and 
interfering angle-dependent Cooper and density-wave tendencies.
The physical justification for using Landau quasiparticles to build our 
theory is the experimental evidence for the existence of the
 Fermi surface in cuprates.\cite{FS_2002}  
Despite  the existence of the Fermi surface, well known is that 
the system does not obey  the laws of  the traditional Fermi liquid
theory.\cite{review_timusk}
This apparent contradiction is what we want to understand theoretically. 
The basic hypothesis is that the normal state is well described by 
the effective action for 
quasiparticles near  Fermi surface with the angle-dependent
two-body interaction $U_l(\theta_1,\theta_2,\theta_3)$, 
and with angle-dependent 
one particle attributes: the quasiparticle weight $Z_l(\theta)$,
the scattering rate $\gamma_l(\theta)$ and the angle-dependent Fermi surface
shift $\mu_l(\theta )$. 
Angle $\theta$ is defined within the so-called $N$-patch model (see inset
on fig.\ref{results}a). In the
limit of low energies this angle
 just parametrizes the position of the particle on
the Fermi curve.
It has been shown by a simple power counting that
only angular dependence 
of the effective interaction
is marginal (or marginally-relevant) 
\cite{Shankar}, and only
terms up to linear in energy are to be kept in the 
renormalization of the angle-dependent selfenergy \cite{Z_EPL}.

As we reduce continuously the energy 
cutoff $\Lambda=8t\exp{(-l)}$, $t$ being the 
nearest neighbour hopping, all mentioned quantities are continuously
renormalized: they evolve in the ``time'' $l$.
Here we employ the Wilson's renormalization scheme, where 
$U_l$ is the effective interaction between electrons {\em within} the ring 
$\pm \Lambda$ around the Fermi surface. This interaction is renormalized 
by  the scattering processes involving all electrons {\em outside} 
the ring $\pm \Lambda$.
Physically more comprehensive interpretation
of the scale $l$ is that it plays the role of the temperature or of the 
experimental probe frequency.

The fundamental difference between our effective action and the effective
action for the traditional Fermi liquid is  that our interaction $U_l$ 
depends on three angles, while in the usual 
Fermi liquid case it depends only on two angles. 
$U_l$ depends on three angles only when for 
any choice of three momenta at the Fermi surface 
 the fourth momentum, by momentum conservation (modulo Umklapp), 
lies also on the Fermi surface. 
Strictly speaking, this is the
case only for 
Fermi curves with flat opposite sheets. 
In this paper we consider the Hubbard model at (nearly) half-filling, where the
Fermi surface is (almost) square so that the above condition is (almost) 
fulfilled.
Namely, systems with the Fermi surface weakly curved by small
 imperfect flatness parameter $\epsilon$ are also  well described 
at all scales $\Lambda > \epsilon$
by the three-angles parametrization.
On the
contrary, in the case of  curved
and non-nested Fermi surfaces (or $\Lambda < \epsilon$) 
only three subsets of the most general
interaction $U_l(\theta _1,\theta _2,\theta_3)$  survive the zero-order
scaling in the limit of low-energies. These subsets are the generalized 
forward and backward amplitudes 
$F_l(\theta ,\theta ')\equiv U_l(\theta ,\theta ',\theta)$ and
$\tilde{F}(\theta ,\theta ')\equiv U_l(\theta ,\theta ',\theta')$, and the
Cooper amplitude $V(\theta ,\theta ')\equiv U_l(\theta ,\theta
+\pi,\theta')$.
That is the familiar Fermi liquid situation \cite{Shankar,ZS}.

\begin{figure} 
\centerline{\psfig{file=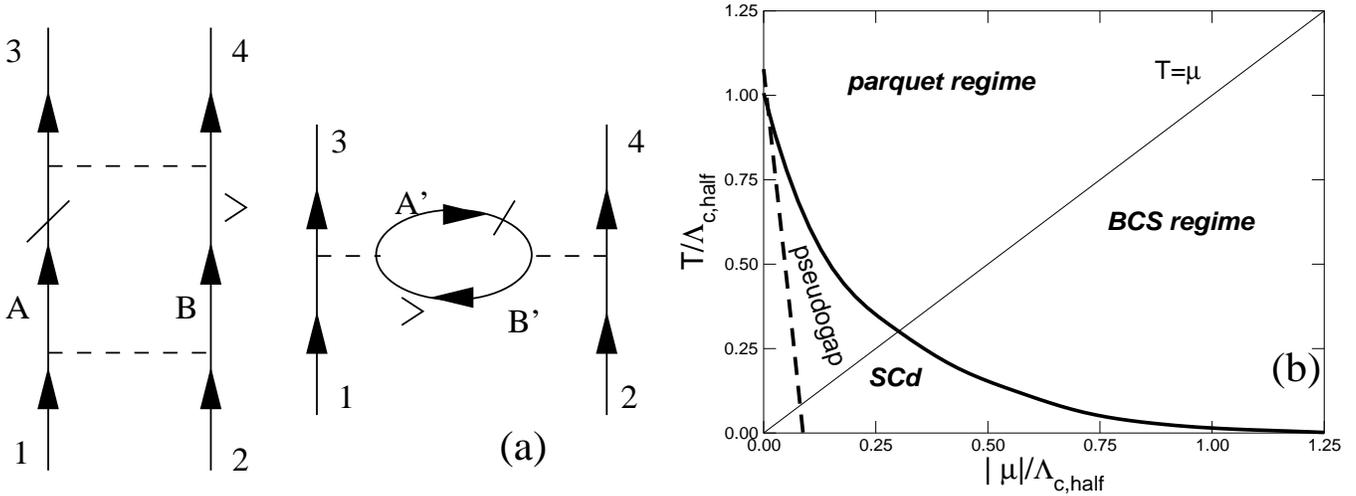,width=0.99\textwidth,angle=0}} 
\hfill 
\caption{\small 
  (a) From the left: Particle-particle (Cooper) and particle-hole 
(density-wave)
    differential bubbles $\beta _{pp}(\theta_1,\theta_2,\theta_3)$ and
$\beta _{ph}(\theta_1,\theta_2,\theta_3)$. (b) The phase diagram. Bold solid
line is the critical scale $\Lambda _c$. In parquet regime $\Lambda_c=T^*$, 
while in BCS regime $\Lambda_c=T_c$. Dashed line is the mean-field critical
temperature for antiferromagnetism. All energies are measured in units
of the critical scale at the half-filling $\Lambda _{c,\mbox{half}}$.} 
\label{bete_FD} 
\end{figure}

Very generally, the one-loop renormalization group equation for $U_l$ has
the following structure
\begin{equation}
\frac{\partial{U_l}}{\partial{l}}
=\beta _{pp}\{ U,
U\} +2{\beta}_{ph}\{ U,U\}-
{\beta}_{ph}\{ U,XU\}-{\beta}_{ph}\{ XU,U\}-X{\beta}_{ph}\{ XU,XU\}.
\label{RG_U}
\end{equation}
One must remember that this is a functional flow equation, i.e. $U_l$ and
all terms on the right-hand side depend on three angles
$(\theta_1,\theta_2,\theta_3)$. Particle-particle (Cooper) and 
particle-hole (density-wave) differential bubbles 
$\beta _{pp}$ and $\beta _{ph}$ are shown on fig.\ref{bete_FD}(a).  
$X$ is the exchange operator defined by $XU(1,2,3)\equiv U(2,1,3)$.
To solve numerically eq.(\ref{RG_U}) we discretize the 
angle $\theta$. The function $U_l(\theta_1,\theta_2,\theta_3)$ is then
represented by a set of coupling constants labeled by three discrete indices.
For the Hubbard model the initial condition is  $U_{l=0}=U_0=$cte.
Upon renormalization $U$ starts to build its angular dependence. 
All coupling constants are found to diverge at the same critical scale $l_c$
like
\begin{equation}
{U}(\theta_1,\theta_2,\theta_3)\rightarrow 
\frac{\tilde{U}(\theta_1,\theta_2,\theta_3)}{(l_c-l)}\; ,
\label{div}
\end{equation}
 where the weights $\tilde{U}$
are model dependent constants. 
This type of solution
is called the fixed-pole solution in contrast to the mobile-pole solution,
where different coupling constants  diverge at different critical scales.
For realistic systems, where the initial coupling is not extremely small, only 
fixed poles are relevant \cite{doucot_00}.

The one-loop renormalization is just a non-prejudiced way of
doing perturbation theory: 
through the Cooper and Peierls loops (fig.\ref{bete_FD}a) all 
{\em two-particles} correlations are taken 
into account.
Contributions involving {\em three- or more-} 
particles correlations are neglegible as 
long as the coupling is weak.
Consequently, one must be careful when
interactions flow towards the strong coupling regime as $l$ approaches $l_c$.
What I stress is that this divergence is not an artifact of the theory, but
the physical reality. In fact,
my very point is that the pseudogap ``phase'' can be seen from its
precursors just like the onset of the BCS superconductivity is seen from 
the metallic phase as a divergence of the effective Cooper amplitude in the 
ladder 
approximation. The difference is that here we calculate the most
general vertex instead of only Cooper amplitude and that we do it by 
the one-loop RG (or, equivalently, parquet) procedure 
instead of the ladder approximation.

The critical scale $l_c$ depends on the bare coupling constant $U_0$ and on the
band filling parametrized by the chemical potential. 
$\Lambda _c=8t\exp{(-l_c)}$ appears to be the
fundamental temperature scale of the  model. The most precise 
non-restrictive interpretation of $\Lambda _c$ is that at this energy
electrons start to build bound states.
The dependence of $\Lambda _c$ on the chemical potential $\mu$ defines 
the phase diagram 
shown on 
fig.\ref{bete_FD}(b). In the BCS regime, the divergence signals the 
onset of  d-wave superconductivity, i.e. $\Lambda _c=T_c$.
Namely, the flow equations reduce to the
RG version of the ladder summation once the nesting contributions to the
flow disappeared at the crossover $T=\mu$.
Concerning the BCS regime I will just remind
that precisely the one-loop RG theory provided the first concrete 
proof for 
superconductivity in the repulsive Hubbard model \cite{ZS}.

In contrast to the BCS regime
where bound states forming at the scale $\Lambda _c$ are simple Cooper
pairs,  in the parquet regime it is much more difficult to characterize the 
two-particle poles
appearing at the energy $\Lambda _c$. We will show that the
divergence of interactions is the onset of the pseudogap.
Before proceeding with theoretical considerations let us call what are the
experimental manifestations of the pseudogap: 
\\(i) At the pseudogap
temperature $T^*$ the Fermi surface starts to be 
progressively destroyed from the van Hove points toward the zone diagonals,
i.e. the pseudogap is angle-dependent, and precisely this angle dependence
depends on temperature. 
\\(ii) Both superconducting (d-wave) {\em and} 
antiferromagnetic 
  {\em short range} correlations are enhanced.
\\(iii) Both magnetic and charge responses are pseudo-gapped.
\\Let's see how our theory reproduces  these three points.

(i) In the parquet regime the available phase space for the selfenergy 
corrections is not restricted \cite{Z_EPL,Honerkamp_self}.
Consequently, the one particle properties can
change radically. Angle-dependent quasiparticle weight renormalizes
according to the flow equation
\begin{equation} \label{RG_polelog}
\partial _l\log{Z}_l(\theta)=\frac{1}{(2\pi)^2}\int d\theta'\; {\cal J}
(\theta',-\Lambda)\eta_l(\theta,\theta')\equiv \eta_l(\theta)\; ,
\end{equation}
${\cal J}
(\theta,\epsilon )$ being the angle dependent density of states at the energy 
$\epsilon$ (measured from the Fermi level).
The quantity $\eta_l(\theta,\theta')$ contains particle-particle (pp) and 
particle-hole (ph) contributions
\begin{equation} \label{eta}
\eta_l(\theta,\theta')\equiv (2X-1)\beta _{pp}\{ U,U\} 
+2\beta _{ph}\{ XU,XU\} +2\beta _{ph}\{ U,U\} -\beta _{ph}\{ U,XU\}-\beta
_{ph}\{ XU,U\} 
\end{equation}
with all terms on the right-hand side taken with external legs
$(\theta_1,\theta_2,\theta_3) =(\theta,\theta',\theta')$.
The interaction inserted in all beta functions obeys
the scaling equation (\ref{RG_U}). 
It can be shown that eq.(\ref{RG_polelog}) reproduces the 
standard Luttinger liquid exponent in the 1D limit \cite{Z_EPL}.
Results for the 2D Hubbard model are on Figure \ref{results}(a). It 
shows the evolution of $Z(\theta)$ as one approaches the
critical scale in the parquet regime. When the interactions start to 
diverge the exponent  $\eta_l(\theta)$ also diverges.
Its angle dependence near the divergence  is shown on fig.\ref{results}(a).
The Fermi surface loss is much faster
near the van Hove points (point A) than on the zone diagonals (point E). 

(ii) To find out which correlations are relevant in the
parquet regime we must allow the theory to choose 
between all possible 2-particle correlations.
For this reason we have to follow the renormalization of several
angle-resolved correlation functions.
The superconducting correlation function $\chi
^{SC}_l(\theta_1,\theta_2)$ measures  correlations between 
the Cooper pairs 
$(\theta_1,\theta_1+\pi)$ and $(\theta_2,\theta_2+\pi)$, all states being at
the Fermi surface. 
The antiferromagnetic correlation function 
$\chi ^{AF}_l(\theta_1,\theta_2)$ correlates two nested electron-hole pairs
$\Psi^{\dagger}_{{\bf k}(\theta_1)}{\bf \sigma}\Psi_{{\bf
k}(\theta_1)+(\pi,\pi)}$ and 
$\Psi^{\dagger}_{{\bf k}(\theta_2)}{\bf \sigma}\Psi_{{\bf k}(\theta_2)
+(\pi,\pi)}$.
The charge density wave correlation function $\chi ^{CDW}_l(\theta_1,\theta_2)$
correlates the nested charge-like electron-hole pairs.
The relevant susceptibility in each channel is the dominant eigenvalue
of the angle-resolved correlation function. 
The corresponding eigenvector
determines the angular dependence of the order parameter.
The scale dependence of the relevant susceptibilities 
  is shown on fig. \ref{results}(b). It should be noted that 
introduction of the selfenergy effects in the renormalization of the
susceptibilities reduces both antiferromagnetic and superconducting 
correlations: they become non-singular and comparable to the $U=0$ case.
The correlation function for $d$-symmetry charge density wave (DDW) 
has  been calculated by
Honerkamp et al.\cite{Honerkamp_DDW}. The corresponding response
is also enhanced but cannot exceed the d-wave superconducting 
susceptibility and
gets weaker with doping  because the DDW is nesting
dependent, just as any charge density wave.

(iii) The behavior of ${\bf q}=0$ 
 susceptibilities in the magnetic and charge sectors
is also very significant. As we approach the critical scale in the
parquet regime the liquid becomes less and less compressible and its
magnetic susceptibility decreases as well \cite{homogeneous_susc}.
The interpretation is straightforward: the system wants to suppress 
the spin and
 charge degrees of freedom from the lowest energies.

The scattering rate $\gamma_l(\theta)$ above the critical scale $\Lambda _c$ 
in the parquet regime was calculated by
Honerkamp \cite{Honerkamp_self} within the  angle-resolved RG.
The result clearly shows a linear temperature dependence for any
position on the Fermi surface. 
Another remarkable one-particle phenomenon is the Pomeranchuk instability of
the Fermi surface. Recent calculations \cite{Pomeranchuk} show 
that the Fermi energy 
renormalization $\mu_l(\theta)$ breaks the lattice symmetry in a way that 
it is shifted
upwards at, e.g. $\theta=0,\pi$ and downwards at $\theta=\pi/2,3\pi/2$. This
is a d-wave chemical potential shift.

In conclusion, all above theoretical results indicate that the rich physics
of the pseudogap in cuprates, as well as the non-Fermi-liquid properties of
the normal phase are reachable by angle-resolved one-loop
renormalization-group upon
the effective action for Landau electrons.

\begin{thebibliography}{9} 
\itemsep 0pt 

 \bibitem{FS_2002}A. Ino {\em et al.} Phys. Rev. B {\bf 65} (2002) 094504.
 \bibitem{review_timusk} T. Timusk and B. Statt, Rep. Prog. Phys. {\bf 62} 
(1999) 61-122.
\bibitem{Shankar} R. Shankar, Rev. Mod. Phys. {\bf 66} (1994) 129-192.
\bibitem{Z_EPL} D. Zanchi, Europhys. Lett. {\bf 55} (2001) 376-382.
\bibitem{doucot_00}F. Vistulo de Abreu and B. Dou\c{c}ot,
Phys. Rev. Lett. {\bf 86} (2001) 2866-2869.
\bibitem{ZS}D. Zanchi and H. J. Schulz, Phys. Rev. B {\bf 54}  (1996) 
9509-9519;\\
Phys. Rev. B {\bf  61}  (2000)
13609-13632; Z. Phys. B {\bf 103} (1997) 339-342.
\bibitem{Honerkamp_DDW} C. Honerkamp {\em et al.}, Eur. Phys. J. B{\bf 27}
 (2002) 127-134.
\bibitem{Honerkamp_self} C. Honerkamp, Eur. Phys. J. B{\bf 21} (2001) 81-91.
\bibitem{homogeneous_susc} D. Zanchi, Ph.D. thesis, 
Universit\'e Paris-Sud (1996) (unpublished); \\C. J. Halboth and W. Metzner, 
Phys. Rev. B {\bf 61} (2000) 7364-7377.
\bibitem{Pomeranchuk} A. Neumayr and W. Metzner, cond-mat/0208431.
\end{thebibliography}
\end{document}